# Hourly U.S.-wide flood simulation beyond the limits of traditional and data-driven models

Wencong Yang[1], Leo Lonzarich[1], Yalan Song[1], Haoyu Ji[1], Ming Pan[2], Kathryn Lawson[1], and Chaopeng Shen[1*]

[1] Civil and Environmental Engineering, The Pennsylvania State University, University Park, PA, USA

[2] Center for Western Weather and Water Extremes, Scripps Institution of Oceanography, University of California San Diego, La Jolla, CA, USA

[*] Corresponding author. cshen@engr.psu.edu

**Abstract**

As increasingly-damaging floods can strike within hours of a storm and in ungauged reaches, hourly network-wide simulation has become critical societal infrastructure. Here we demonstrate a multi-timescale physics-embedded learning model which outperforms the United States' operational system and surpasses AI-based systems at flood peaks. Covering more than 800,000 river reaches of the conterminous U.S., the model elevates median hourly Nash-Sutcliffe efficiency at 2,831 gauges to 0.683 from 0.461 for the operational National Water Model v3.0, and narrows flood-peak timing errors from 7-8 hours to 4.5-6 hours. δHBV2.0MTS-MC captures 33% more ≥50-year floods than NWM3.0 and 159% more than an operational LSTM baseline. Against recent AI models, overall hourly skill is comparable while rare-flood accuracy is distinctly higher, with relative peak-magnitude error reduced by 34% for ≥100-year floods. It combines long-term hydrologic context, short-term shocks, and infiltration excess to resolve extraordinary hourly peaks not visible on a daily plot. Process-model parameters and hourly discharge are produced for every reach, seamlessly covering the continent at 7.2 $km^2$ median resolution. This candidate for the next-generation National Water Model sets a new operational accuracy level for national-scale flood prediction.

## Introduction

Floods are among the deadliest and costliest natural hazards[1], and the events that cause the most damage develop within hours of a storm, often in reaches that are not monitored[2]. The July 2025 Texas Hill Country flood was caused by intense, concentrated overnight rainfall that drove the Guadalupe River from low flow to a destructive crest within hours, in small, steep basins in a stream branch that was not monitored[3]. Flooding in the Carolinas repeated the pattern across headwater catchments[4], so neither the timing nor the location of the worst response was directly observed. In these events resolving the hourly peak is much more important than the mean state. Predictions useful for warnings must therefore be accurate at flood peaks and high temporal frequency, reliable even for rare events, and available along the entire river network rather than at fixed gauges alone.

Operational hydrologic forecasting and water-management systems require hourly streamflow simulations over spatially-continuous continental-scale river networks. Sub-daily resolution is needed because flood hazards are often concentrated as short-duration peak flows, which can be substantially smoothed or underestimated in daily streamflow records. Applications such as flash-flood forecasting[5], reservoir operations[6], and flood-risk assessment[7] depend on accurate prediction of both the precise timing and location of hydrologic responses. At the same time, continental-scale deployment is necessary to provide consistent predictions across millions of interconnected river reaches, including ungauged and sparsely gauged locations[8]. These demands require hydrologic models that can resolve hourly-scale dynamics while maintaining long-term temporal consistency and seamless spatial coverage over large areas such as the conterminous United States (CONUS).

Traditionally, sub-daily distributed process-based models have provided the main pathway for physically-consistent streamflow simulation at high temporal resolution across large river networks[8–10]. The National Water Model (NWM) applies the WRF-Hydro model for hourly continental streamflow simulation across the United States[8]. Global reach-level flood reanalysis work has coupled VIC runoff generation with RAPID routing to produce 3-hour discharge for millions of river reaches[10]. These systems demonstrate the value of high-resolution distributed physical simulations, but parameter estimation for each model component and location remains challenging at sub-daily continental scales. Short time steps increase computational cost, while spatial heterogeneity across large domains limits the transferability of conventional calibration strategies.

Machine learning techniques have recently advanced both hourly streamflow prediction and distributed hydrologic modeling. Multi-timescale long short-term memory (MTS-LSTM) uses a recurrent model to combine long histories at coarser temporal resolution with recent inputs at finer resolution, reducing the burden of long hourly warmup while preserving information from antecedent conditions[11]. Multi-frequency LSTM (MF-LSTM) further generalizes this idea by allowing one LSTM cell to process different temporal frequencies and different input dimensions in a shared architecture for hourly streamflow prediction[12]. H-Diffusion uses a denoising diffusion model to generate probabilistic hourly hydrographs and supports data assimilation[13]. These models show the strength of data-driven multi-timescale learning and the high predictive skill already achieved by modern neural architectures[11,14,15], but they are spatially lumped, black-box predictors that do not simulate physically-meaningful internal variables for multivariate learning. They may also face challenges when extrapolating to unseen extreme events[16–19] or when applied in regions with limited streamflow observations[20–22].

Distributed data-driven or hybrid models have demonstrated high performance, but they are mostly daily and cannot capture the higher casualty-inducing peaks that are visible at the hourly scale but not at the daily scale. Pretrained LSTM runoff predictors have been applied over grids in LSTM-RAPID[23] or over subbasins in spatial recursive modeling[24], DROP[25], LSTM-LTI[26], and LSTM-TRoute[27], with runoff subsequently routed through river networks. Others learn river-network routing directly[28], or learn both runoff and routing end to end, as in RiverMamba[29] and grid-based networks[30]. Differentiable or hybrid modeling instead preserves process structure, embedding hydrologic equations in automatic-differentiation frameworks[17,31–35]. However, existing distributed versions run at daily time steps only[33,36–38]. Song et al.[39], Bindas et al.[40], and Ji et al.[41] combined the Hydrologiska Byråns Vattenbalansavdelning (HBV) model with Muskingum-Cunge (MC) routing and learned spatially continuous parameters, but this model series operates only at the daily scale and mainly describes saturation excess. Not only does training at the hourly scale pose a daunting computational challenge for the already large training tasks, but the process representation may also be operationally insufficient for hourly floods, i.e., lacking infiltration excess[42,43]. Furthermore, it is uncertain how to best balance the effects of long-term water storage and statistically-rare hourly events in training.

Here we demonstrate a new skill level in high-quality, high-resolution (~7.2 km$^2$) hourly flood prediction, configured for seamless national-scale forecasting platforms as a next-generation critical infrastructure. This system contains a multi-timescale structure that preserves long antecedent hydrologic states, a rapid runoff formulation for hourly-scale responses, and coupled Muskingum-Cunge river routing, in an end-to-end differentiable system trainable on big data. This system, denoted δHBV2.0MTS-MC, offers distinctly higher hourly accuracy than the operational national models as well as purely AI-based systems, drastically reduces peak timing and magnitude errors, and captures many more rarest flood events.

## Results

### *A new skill level for national-scale flood simulation*

Across 2,831 CONUS gauges, the multi-timescale differentiable HBV models cleanly exceed the extensively calibrated National Water Model (NWM3.0) and LSTM-TRoute in aggregate hourly accuracy (Fig. 1). δHBV2.0MTS-MC achieves a median hourly Nash-Sutcliffe Efficiency (NSE) of 0.683 (vs. 0.461 for NWM3.0), a median Kling-Gupta Efficiency (KGE) of 0.719 (vs. 0.591), and a median absolute percent bias (|PBIAS|) of 10.24% (vs. 15.28%) (Fig. 1). As another operational baseline, LSTM-TRoute within the current Water Resources Modeling Framework[27] achieves a relatively low median hourly NSE of 0.359. Similarly, for both the high-flow (|FHV|) and low-flow (|FLV|) ends of the distribution, the improvements above NWM3.0 are obvious with a 20%-21% relative reduction in median errors. As will be discussed below, between the two routing variants of our framework, Muskingum-Cunge is the operationally-relevant configuration. While NWM3.0 relies on basin-specific calibration with donor regionalization (which involves many ambiguous choices), δHBV2.0MTS-MC discovers continent-shared attribute-to-parameter relationships automatically, creating a new operational accuracy level for national-scale flood forecasting.

While CONUS gauges compare continentally-scale distributed models, differentiable models' advantages recur over the smaller, reference basins in the CAMELS dataset, which allows for uniform-setup benchmarks against established data-driven models, using realizations released by their original teams. δHBV2.0MTS-MC (hourly median NSE ~0.710) and its unit-hydrograph variant, δHBV2.0MTS-UH (~0.716) both surpass NWM3.0's hourly NSE of 0.545, while reducing median |FHV| by a relative 22%-25% (Table 1). While differentiable models' NSE and KGE values are comparable to recent data-driven hourly benchmarks, their high-flow biases are distinctly lower than those of MTS-LSTM and MF-LSTM, showing the process-based modules' apparent robustness for peaks. For these smaller basins, this advantage primarily arises from the behavior of mass-conservative runoff generation under less frequently-seen magnitudes rather than the distributed routing scheme.

The benefit of explicit river-network routing emerges primarily at larger basin scales. While δHBV2.0MTS-MC skill remains relatively stable across basin sizes (Supplementary Fig. S4), above 10,000 $km^2$, unit-hydrograph routing degrades at larger basins, highlighting the effects of nonlinear flood dynamics (Supplementary Fig. S2). δHBV2.0MTS-MC achieves a median NSE of 0.668 (versus 0.546 for δHBV2.0MTS-UH), and a median KGE of 0.736 (versus 0.529). More importantly, it propagates discharge through the connected river network and therefore produces hourly streamflow at every reach, whereas the unit-hydrograph configuration is designed primarily for predefined gauged basin outlets.

Model skill varies systematically with climate and terrain: median hourly NSE falls from 0.729 in humid basins (aridity index 0.2-0.8) to 0.343 in the most arid ones (1.5-10), and from 0.696 to 0.552 as snowfall fraction rises from below 0.05 to 0.40-0.60 (Supplementary Fig. S4). The bias reduction relative to NWM3.0 is concentrated in the arid West and the Great Plains (Supplementary Fig. S3). Unsurprisingly, the national-scale gain is broad but heterogeneous, with arid and snow-dominated hydrology remaining the most challenging regimes[8,44,45].

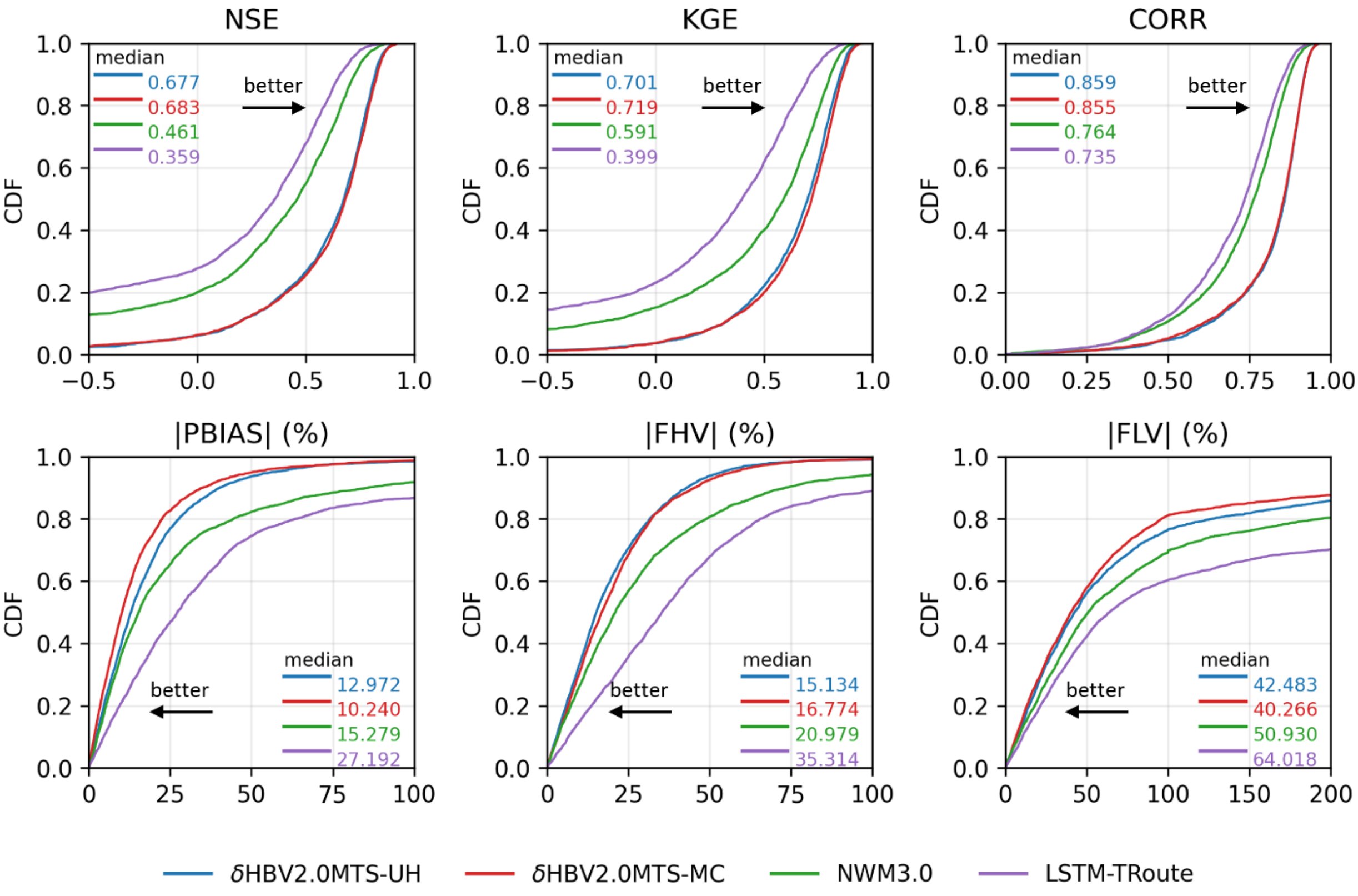

*Fig. 1 | Cumulative distribution functions (CDFs) of Nash-Sutcliffe efficiency (NSE), Kling-Gupta efficiency (KGE), Correlation (Corr), absolute percent bias (|PBIAS|), absolute high-flow bias (|FHV|), and absolute low-flow bias (|FLV|) for δHBV2.0MTS-MC (primary multi-timescale model with Muskingum-Cunge routing),* δHBV*2.0MTS-UH (corresponding unit-hydrograph variant), NWM3.0 (National Water Model version 3.0)[8], and LSTM-TRoute[27] (a purely data-driven model currently available) over conterminous US gauges during the 2009–2018 test period.*

***Long memory, fast runoff, and dynamical routing for rare floods***

Critical to the core mission, big-data trained models capture substantively more rare peaks than either NWM3.0 and LSTM-Troute, compressing the risks of missing flood events. Under the criteria of the absolute percentage error (APE) ≤20% and timing error (TE) ≤3 h, δHBV2.0MTS-MC captures 33% more than NWM3.0 and 159% more than LSTM-Troute (Fig. 2c). This is a much tighter TE criteria compared to Nearing et al.[15] which is 2 days. Relaxing the APE and TE criteria (30%, 5h) leads to more captured events but δHBV2.0MTS-MC similarly captures ~20% more events than NWM3.0 and 100% more than LSTM-Troute (Supplementary Fig. S5). This contrast suggests δHBV2.0MTS-MC is more robust for the more extreme events. NWM3.0 in fact captures more flood peaks than LSTM-TRoute, attesting to the weakness of purely data-driven networks and the value of process representation. However, the median timing error is 7-8 hours, compared to the 4.5–6 h for δHBV2.0MTS-MC (Fig. 2a), presumably because big-data training better propagates information about floodwave propagation to the flood routing parameters. The same reduction in error can be witnessed on the CAMELS dataset from the lower |FHV| compared to both MTS-LSTM and MF-LSTM (Table 1).

Improved descriptions of floods arise from δHBV2.0MTS-MC's lower errors with flood peak magnitude and timing, especially the rarest events, which can heavily degrade purely data-driven models. Across CONUS, δHBV2.0MTS-MC showed significantly lower APE and lower TE in all classes (Fig. 2a) than NWM3.0 and LSTM-TRoute across different return-period classes. One lever against the HBV models is the lack of descriptions for reservoir operations, which are indeed represented in NWM3.0, so future improvements can target this gap, but this factor is not enough to diminish differentiable models' advantages. On the well-benchmarked CAMELS basins, the lumped MTS-LSTM shows an APE of 59.4% and a median TE of 5.5 hours for >100-year floods (Fig. 2a), which is the reason why it misses many more events; whereas δHBV2.0MTS-MC reduces peak-magnitude error relatively by 34% and lowers the median timing error to 3 h.

Hourly resolution itself is a prerequisite for reproducing the short, high-amplitude peaks that are averaged down at the daily scale, but hourly floods are harder to learn from because they occupy a smaller representation in the training dataset. For observed hourly peaks above 1 mm $h^{-1}$, the hourly model reduces median APE by 36–42% relative to the daily model δHBV2.0-MC (Fig. 2b). Because hourly flood magnitude and timing cannot be evaluated directly from daily predictions, external benchmarking is restricted to models with native hourly outputs; the daily δHBV2.0-MC is included specifically to isolate the effect of temporal resolution within the same model family. However, at the hourly scale, a challenge facing data driven models is that flood peaks can be abrupt and very rare, challenging stand-alone LSTMs of various kinds. Our comparison adds to the growing evidence that process-based and hybrid hydrologic models can be more robust than stand-alone LSTMs when predicting extreme or unseen conditions. Previous studies have attributed this to physically constrained responses[16,19,31,46] and LSTM's state saturation limitation[17]. At hourly resolution we also need to resolve routing, as explained below.

Hourly spatiotemporal resolutions alone, however, are insufficient – long antecedent memory, rapid runoff response, and dynamic river routing provide complementary improvements needed to reach the final performance.

- First, groundwater and long-term storage memory control available source-area pore spaces and modestly impact peak magnitude in regions with high storage-streamflow connections. This long-term effect is quantified by the multi-timescale δHBV2.0MTS-MC's improvement in median NSE from 0.643 to 0.710 and reduction in |FLV| from 69.27% to 51.37% (Table 1), compared with the independently trained hourly-only δHBV2.0h-MC.
- Second, compared with δHBV2.0MTS-MC-base, which lacks the fast infiltration-excess module, δHBV2.0MTS-MC leaves aggregate metrics essentially unchanged (Table 1) while lowering flood-peak errors across all return-period classes (Supplementary Fig. S6a). The largest reductions occur for peakier events in smaller and drier basins (Supplementary Fig. S6b and Supplementary Text D), consistent with the intended role of fast runoff generation. On a side note, this contrast also showcases why the aggregate metrics are not reliable measures of peak flow success.
- Third, floodwave propagation is nonlinear and is accelerated during high flows, and MC's flow-dependent celerity significantly better resolves such effects than a learnable but fixed unit hydrograph (UH). This results in a greater extent of timing error (TE) compression for the peakier floods, and the reduction of median TE can be as large as a relative 50% for the peakiest events (Supplementary Fig. S7b). In terms of catchment area, the gap between the dynamical routing version and the UH is not significant on the smaller CAMELS basins (Table 1), where the UH version could sometimes be better, but can reduce the APE from median 40% to 32.5% on CONUS basins exceeding 10,000 $km^2$ (Fig. 1 & Supplementary Fig. S7a). Apparently both are performant, but the dynamical routing version (MC) is more suitable for modeling flashier floods and large stem riverine flooding.

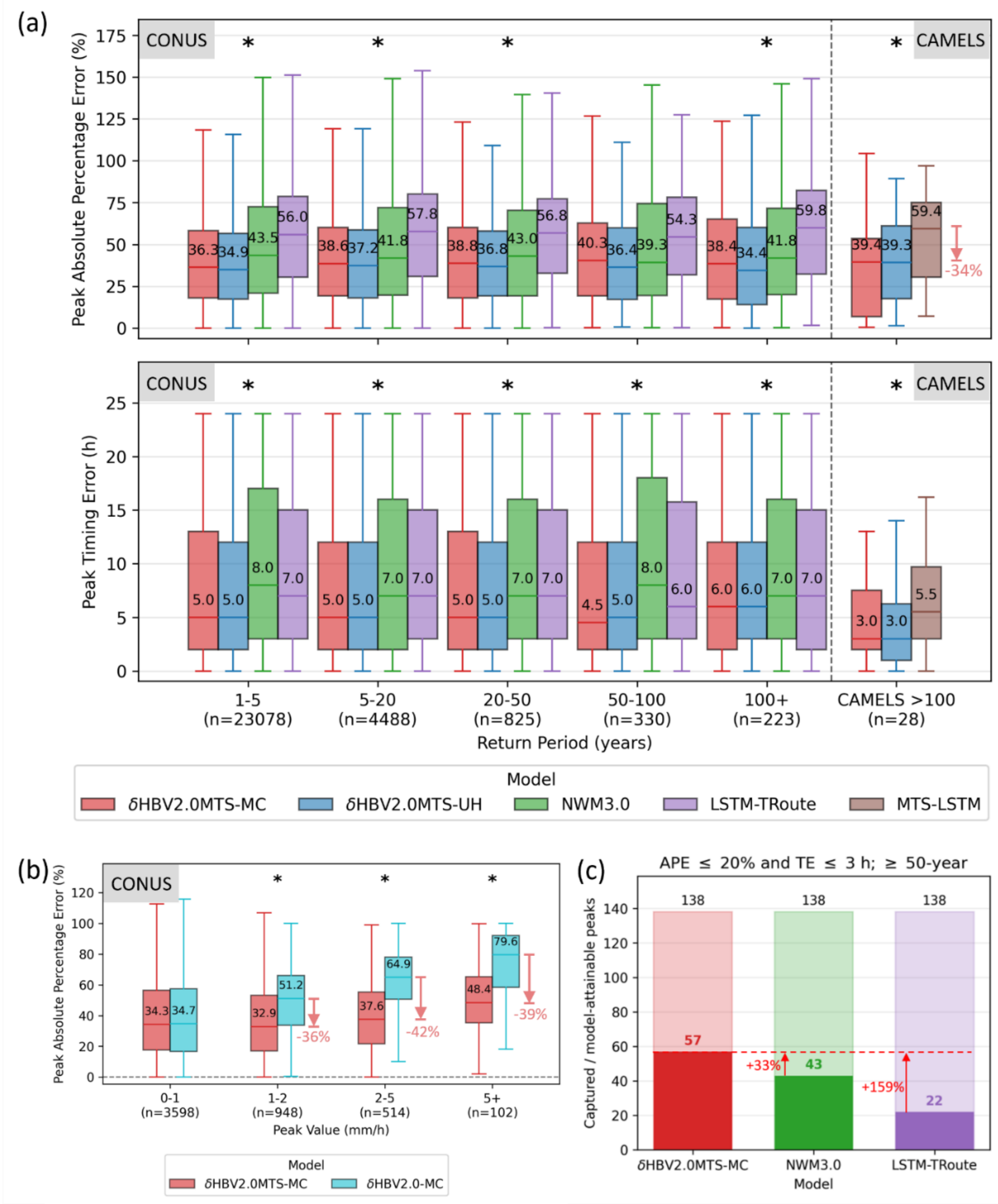

*Fig. 2 | a, Absolute percentage error (APE) and timing error (TE) for hourly flood peaks by return-period class in conterminous US and CAMELS basins, 2009–2018, for δHBV2.0MTS-MC (primary model), δHBV2.0MTS-UH, NWM3.0[6], LSTM-TRoute[27] and MTS-LSTM[11]. b, APE by observed hourly peak-flow range, 2009–2010, for δHBV2.0MTS-MC and the daily δHBV2.0-MC, whose daily values are repeated over 24 h; peaks exceed the 1-year return period. c, Number of model-attainable extreme flood peaks captured in conterminous US basins, defined as events captured by at least one evaluated model with APE ≤ 20% and TE ≤ 3 h. In a and b, asterisks denote significantly lower error for δHBV2.0MTS-MC (paired Wilcoxon signed-rank tests[47]), against NWM3.0 and MTS-LSTM in a and against δHBV2.0-MC in b. Boxes, 25th–75th percentiles; centre line, median; whiskers, most extreme values within 1.5× the interquartile range.*

*Table 1 | Median metrics for δHBV2.0MTS-MC, δHBV2.0MTS-UH, δHBV2.0MTS-MC-base, δHBV2.0h-MC, NWM3.0[8], MF-LSTM[12], and MTS-LSTM[11] on CAMELS basins during the 2009–2018 test period. For MTS-LSTM and MF-LSTM, we use the mean performance of individual realizations rather than the reported ensemble metrics, ensuring a fair one-to-one comparison with the single-realization δHBV2.0 model families. Here δHBV2.0MTS-MC, δHBV2.0MTS-UH, δHBV2.0MTS-MC-base, and δHBV2.0h-MC were trained exclusively on the CAMELS basins using data from 1991–2003, consistent with MF-LSTM and MTS-LSTM. All δHBV2.0 variants, NWM3.0, and LSTM-TRoute use Analysis of Record for Calibration (AORC) forcing[48]; whereas MTS-LSTM and MF-LSTM use North American Land Data Assimilation System (NLDAS) forcings[49] from their original studies.*

| Model | Description | NSE | KGE | Corr | \|PBIAS\| % | \|FHV\| % | \|FLV\| % | Reference |
|---|---|---|---|---|---|---|---|---|
| δHBV2.0MTS-MC | Primary model. Distributed multi-timescale HBV with daily state warm-up, hourly saturation-excess and infiltration-excess processes, and Muskingum-Cunge routing over the Hydrofabric river network; trained end to end. | 0.710 | 0.736 | 0.864 | 12.408 | 13.370 | 51.370 | This study |
| δHBV2.0MTS-UH | Replaces Muskingum-Cunge routing with learned distributed unit hydrographs while retaining the multi-timescale HBV runoff-generation model. | 0.716 | 0.758 | 0.867 | 10.841 | 12.672 | 49.752 | This study |
| δHBV2.0MTS-MC-base | Ablation of the primary model without the expanded fast-release parameter range or infiltration-excess module. | 0.706 | 0.744 | 0.858 | 10.597 | 14.049 | 50.180 | This study |
| δHBV2.0h-MC | Ablation in which the hourly-only HBV-MC model is independently trained with a 1,440-hour warm-up; used to assess the benefit of the multi-timescale daily-to-hourly warm-up. | 0.643 | 0.700 | 0.830 | 13.298 | 13.944 | 69.272 | This study |
| NWM3.0 | Operational, distributed continental-scale hydrologic model based on WRF-Hydro. | 0.545 | 0.684 | 0.792 | 11.892 | 17.050 | 43.244 | Cosgrove et al.[8] |
| LSTM-TRoute | Operational distributed hourly runoff model using basin-trained LSTMs, coupled with t-route for river-network routing. | 0.457 | 0.410 | 0.766 | 27.984 | 37.904 | 66.660 | Patel et al.[27] and NOAA-OWP[50] |
| MTS-LSTM | Lumped data-driven model that combines long antecedent histories at a coarse temporal resolution with recent inputs at a finer resolution. | 0.726 | 0.723 | 0.869 | 12.511 | 16.789 | 37.557 | Gauch, Kratzert, et al.[11] |

| Model | Description | NSE | KGE | Corr | \|PBIAS\| % | \|FHV\| % | \|FLV\| % | Reference |
|---|---|---|---|---|---|---|---|---|
| MF-LSTM | Lumped data-driven model that processes multiple temporal frequencies and input dimensions using a shared LSTM cell. | 0.716 | 0.732 | 0.863 | 12.845 | 15.364 | 32.411 | Acuña Espinoza et al.[12] |

***Flood events up close: skill across diverse flood dynamics***

For some large hourly peaks on representative sites (Fig. 3a,b,d), δHBV2.0MTS-MC can finally resolve the magnitude and rapid responses while NWM3.0, MTS-LSTM and the ablated δHBV versions (Supplementary Fig. S6) failed to reproduce these peaks. Particularly, in some small headwater basins (Fig. 3a and d), MTS-LSTM severely underestimates the peaks and could not rise fast enough due to the unprecedented nature of these peaks, and the δHBV2.0MTS-MC-base also generates too little flood volume, likely because infiltration excess strongly activates under such extreme storms. The improvement is not limited to narrow flash peaks: the model also closely reproduces a broad, long-duration flood in a medium-sized basin (Fig. 3c) and the magnitude and timing of a multi-peak event (Fig. 3d), where MTS-LSTM still vastly underestimates. The underestimation δHBV2.0MTS-MC-base is less severe in Fig. 3c than in Fig. 3a,3b,3d, showing that the infiltration excess is most important in smaller catchments. Hence, while the infiltration excess module may not show strong benefits in aggregate statistics across hundreds of basins, in some situations it is a must.

The examples also illustrate the importance of antecedent storage from long-term hydrologic inputs and explicit river routing under different hydrologic conditions. During the snowmelt season in snow-dominated mountainous basins, δHBV2.0MTS-MC reproduces the observed diurnal streamflow cycles more closely than the hourly-only δHBV2.0h-MC (Fig. 3e–f), which does not account for long-term (>2 months) storage and accumulated snowpack. This initial condition cannot be reproduced without knowing snow accumulation that occurred throughout the months. In the two large basins (>11,000 $km^2$; Fig. 3g–h), the Unit Hydrograph version (δHBV2.0MTS-UH) produced floodwaves that are too slow and peaks that are too low. Here, explicit reach-by-reach routing resolves sub-daily flood-wave translation through long river networks. Flood convergence could accelerate the arrival of peaks due to nonlinearly forcing a higher flow celerity is visibly causing the multi-hour lag, which would not appear in analysis at the daily scale. Nevertheless, the few hours are critical for flood warnings.

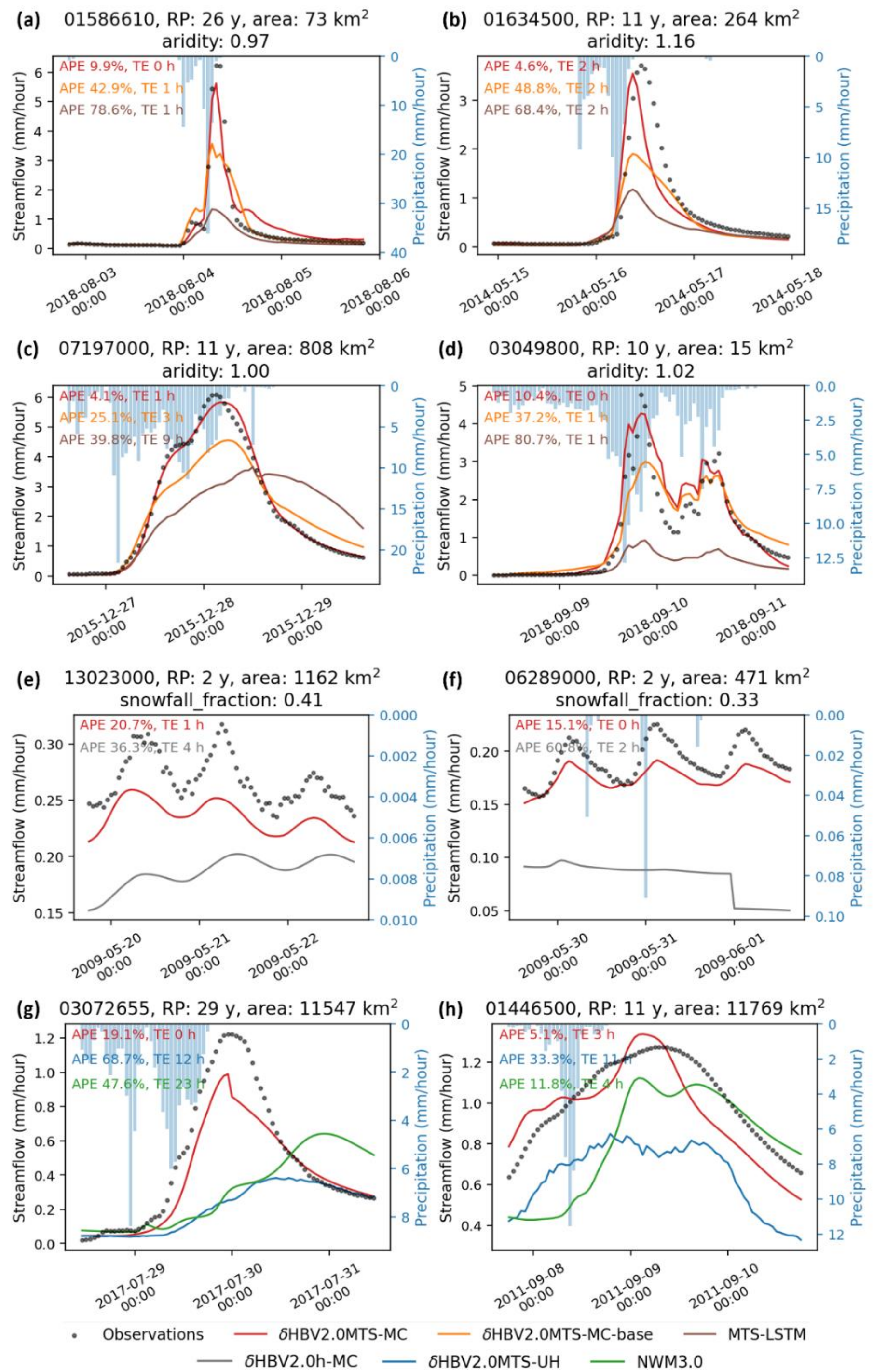

*Fig. 3 | Typical example observed and simulated flood hydrographs for (a-d) large peaks, (e-f) storage-driven floods, and (g-h) floods in large basins. Models include δHBV2.0MTS-MC (primary model), δHBV2.0MTS-MC-base (without the hourly fast-runoff formulation), δHBV2.0h-MC (hourly-only variant), δHBV2.0MTS-UH (unit-hydrograph variant), NWM3.0[8] and MTS-LSTM[11]. Absolute percentage error (APE) and timing error (TE) for flood peaks are reported for each model. RP: return period. Some typical hydrographs were selected.*

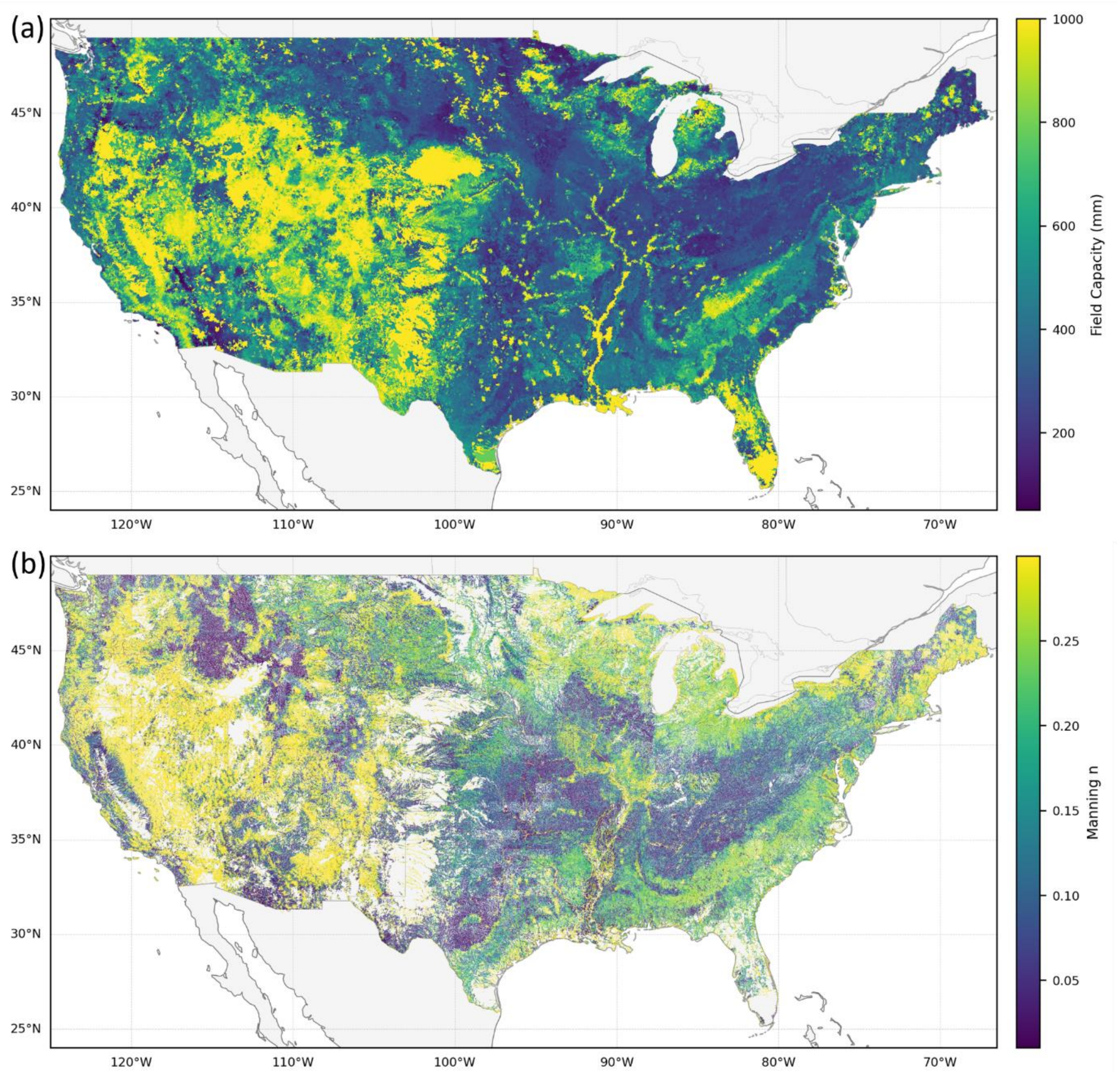

*Fig. 4 | Learned parameter examples for δHBV2.0MTS-MC (primary multi-timescale model with Muskingum-Cunge routing): (a) HBV field capacity ($\theta_{FC}$) and (b) routing roughness parameter (Manning's n).*

## Discussion

### *Interpretable parameters and hourly discharge for every reach*

As an unfamiliar feature, the δHBV2.0MTS-MC model, one trained, effortlessly produces spatially continuous parameter fields over the entire ~7.2 km$^2$ Hydrofabric unit basins and river reaches (Fig. 4). The learned fields include runoff-generation parameters such as HBV field capacity (Fig. 4a) and hourly-relevant routing parameters such as Manning's roughness (Fig. 4b), which are not directly measured or are only sparsely available. Lumped data-driven predictors offer no equivalent, since they emit discharge at trained gauge locations without internal physical states or transferable parameters. Previously, obtaining such parameter fields would have taken years (sometimes decades) of repeated calibration, uncertain regionalization techniques and extensive verification efforts, and when a model is evolved structurally these efforts are needed again. However, a trained model now can produce these fields under a few minutes for the entire continent, with strong generalization performance. It also sets up the stage for smart ensembles, probabilistic forecasting[51], and data assimilation[13,52].

The learned parameter show physically coherent continental patterns. HBV field capacity is larger in the seasonally dry interior West and lower in the humid East (Fig. 4a), consistent with greater root-zone storage under aridity and precipitation seasonality[53]. Manning's $n$ is higher in mountainous/headwater regions and lower in lowlands and coastal areas (Fig. 4b), consistent with CONUS channel parameterizations[54,55]. These patterns support, though do not directly validate, the physical meaning of the learned parameters.

***From lumped to high-resolution: A capability change in seamless flood detection***

δHBV2.0MTS-MC is more computationally demanding during both training and test, but it enables flood detection in ungauged and unmonitored reaches. These spatially complete parameter fields support high-resolution, spatially distributed simulation over the river network at hourly scale. MTS-LSTM and MF-LSTM primarily provide predictions at individual gauges ("lumped"), whereas δHBV2.0MTS-MC generates runoff throughout the flow network and explicitly propagates it through connected river reaches. Because the model operates directly on the NextGen Hydrofabric of the National Water Model, its learned parameters and hourly discharge are consistent between training and deployment and are natively deployable in the same operational infrastructure.

Here we show that detailed accounting and learning of critical high-resolution hydrologic processes is plausible at high spatiotemporal resolution, but it requires scale-specific enhancements in both time and space. Early sections showed that at the hourly scale long antecedent memory, sufficiently fast runoff generation mechanisms and dynamical routing are all needed to optimally capture floods. These necessities impose daunting computational and operational challenges (to training over this much data at high-resolution, we have nearly 80,000 HBV models running concurrently) which for the first time have been shown to be addressable using a hybrid learning approach.

A societally-relevant consequence of big-data training is also the collapse in the cost of continental-scale parameterization and simulation, which is crucial for democratizing the information: the entire CONUS model is trained on a single end-to-end training run — about five days on one A100 GPU for the unit-hydrograph configuration — and a 10-year hourly simulation of the full network of more than 800,000 reaches with exact Muskingum-Cunge routing takes about four hours on a single GPU. This a substantial change in efficiency enables faster real-time flood warnings at higher resolutions, larger coverage, and higher operational readiness.

***Limitations***

Several limitations remain. First, the training domain excludes gauges larger than 50,000 $km^2$ due to computational burden, so performance on the largest river systems remains untested. Extending the framework to large basins will also require attention to reservoir operations, diversions, water use, floodplain storage, and long routing memory. Second, skill remains low in arid, steep, and snow-dominated basins, likely reflecting both forcing uncertainty and simplified process representations, particularly for snow and infiltration. We thus see the forcing quality as a core difficulty to overcome in the future. Third, spatially complete simulation is demonstrated as a capability rather than validated at ungauged reaches; establishing accuracy there requires held-out-gauge experiments.

## Conclusions

We demonstrate a continental-scale hourly hydrologic model that combines multi-timescale HBV runoff generation, rapid saturation- and infiltration-excess responses, and explicit Muskingum-Cunge river routing in an end-to-end high-resolution training framework. The model sets a higher skill level for operationally relevant continental simulation, raising median hourly NSE across 2,831 CONUS gauges from 0.461 for NWM3.0 to 0.683 while also improving flood-peak magnitude and timing. It captures 33% more peaks than NWM3.0 and 159% more than the operational LSTM-TRoute baseline. Against research-level data-driven models, it reaches comparable overall skill but becomes distinctly more accurate toward rare floods, reducing peak-magnitude error relatively by 34% for >100-year events compared with MTS-LSTM. Controlled comparisons further show why these gains emerge: multi-timescale initialization preserves long antecedent states, rapid runoff processes improve short and intense responses, and explicit river routing reduces the increased errors as network scale grows.

## Methods

The framework developed here is a differentiable distributed runoff-routing model for hourly streamflow simulation over spatial units defined by a basin delineation and hydrographic network dataset. It combines a multi-timescale (MTS) HBV runoff-generation module with either distributed unit-hydrograph (UH)[56] or Muskingum-Cunge[57,58] streamflow routing. Building on Song et al.[39], we extend its daily runoff-generation and separately trained routing system to hourly simulation through a daily-to-hourly multi-timescale HBV cascade, hourly fast-runoff modifications, and joint end-to-end training of runoff generation and MC routing.

### *Architecture*

The overall model can be written concisely as

$$\theta = f_H(X, A_B), \qquad \phi = f_U(A_L), \qquad R = H(X;\, \theta), \qquad \hat{Q} = U(R;\, \phi, G) \tag{1}$$

where $X$ denotes meteorological forcings, $A_B$ and $A_L$ respectively denote basin ($B$) and reach or path ($L$) attributes, $\theta$ denotes HBV runoff-generation parameters produced by neural network $f_H$, $\phi$ denotes routing parameters produced by neural network $f_U$, $R$ is unit-basin runoff, $G$ is the spatial connectivity information, and $\hat{Q}$ is routed streamflow at gauges or reaches. The parameter neural networks output values in (0,1), and each physical parameter is transformed to its prescribed range before it enters the process equations. This bounded parameterization preserves the interpretability of the process model while allowing parameters to vary spatially and temporally.

The architecture has three stages (Supplementary Fig. S1). First, a daily HBV model is run over a daily warmup window to initialize snow, meltwater, soil moisture, upper-zone, and lower-zone states. Second, the initialized states are transferred directly to an hourly HBV model, which simulates local runoff over the hourly prediction window. Third, routed streamflow is computed from the local runoff using either the distributed UH router or the MC router. All components are implemented in a PyTorch-compatible differentiable format, so gradients from streamflow losses at gauges can be propagated through routing, runoff generation, and the neural networks that estimate model parameters. During training, the loss is applied after routing, so runoff-generation and routing parameters are learned jointly from downstream streamflow observations.

***Runoff generation: distributed multi-timescale HBV***

Runoff generation is based on δHBV2.0[39], a conceptual hydrologic model with snow, soil-water, and groundwater reservoirs. For each unit basin $i$, the HBV state vector is

$$S_i^{(t)} = \left[S_{p,i}^{(t)},\ S_{melt,i}^{(t)},\ S_{s,i}^{(t)},\ S_{UZ,i}^{(t)},\ S_{LZ,i}^{(t)}\right] \tag{2}$$

where $S_p$ is snowpack, $S_{melt}$ is meltwater retained in the snowpack, $S_s$ is soil moisture, $S_{UZ}$ is upper-zone storage, and $S_{LZ}$ is lower-zone storage. At time step $t$, HBV updates these states from precipitation, temperature, potential evapotranspiration, and the HBV parameters θ.

Following the multi-timescale modeling structure of Gauch et al.[11], state initialization and prediction are handled at different temporal resolutions. Specifically, the daily model runs for a 351-day warm-up period, followed by a 336-hour simulation with the hourly model. This design allows a long antecedent period to be summarized efficiently at the daily scale while preserving hourly runoff dynamics during the training and evaluation window.

Specifically, the daily HBV model first maps the daily forcing sequence to the final warmup state,

$$s_{i,h}^{(0)} = H_d\left(X_{i,d}^{(1:T_d)}; \theta_{i,d}\right) \tag{3}$$

where $H_d$ is the daily ($d$) HBV model and $s_{i,h}^{(0)}$ is the initial state (0) for hourly ($h$) simulation. The final daily states (at time $T_d$) are used directly as the initial hourly states. The hourly model $H_h$ then computes

$$R_i^{(1:T_h)} = H_h\left(X_{i,h}^{(1:T_h)}, s_{i,h}^{(0)};\ \theta_{i,h}\right) \tag{4}$$

At sub-daily scales, however, runoff generation is strongly influenced by short-duration rainfall intensity and rapid runoff release, whereas daily process representations tend to smooth these short, intense responses[59]. We therefore introduce a rapid saturation-excess runoff formulation and an infiltration-excess runoff module[60].

We refer to the resulting models as δHBV2.0MTS. We also evaluate a baseline that excludes the rapid saturation-excess and infiltration-excess module; when coupled with MC routing, this configuration is denoted δHBV2.0MTS-MC-base.

***Routing: unit hydrographs and Muskingum-Cunge method***

The model supports differentiable UH and MC routing options, as Supplementary Fig. S1 shows. The UH router provides a computationally efficient, conceptual representation of aggregated basin-to-gauge travel through convolution, whereas the MC router explicitly propagates discharge through the river network using a more physically based and computationally intensive formulation, allowing us to evaluate the performance gain from increased routing complexity.

***Experimental setup***

*Datasets*

The experiments use (1) CAMELS[61] and (2) the same CONUS training gauges as Song et al.[39], with both datasets represented on the Hydrofabric river-network discretization[8,62]. The full Hydrofabric domain contains more than 800,000 unit basins and associated river reaches, with a median unit-basin area of 7.2 $km^2$ and a median reach length of 4.6 km. Hydrofabric also provides the directed upstream-to-downstream topology used by the distributed routing modules. Gauges are retained only when the reported gauge drainage area is consistent with the Hydrofabric accumulated unit-basin area, using a 15% relative area-difference threshold. We also exclude gauges with drainage area greater than 50,000 $km^2$ to keep the training problem focused on basins with tractable contributing-network sizes. The resulting datasets contain 492 CAMELS gauges with 27,177 contributing Hydrofabric unit basins and 2,831 CONUS gauges with 390,777 contributing unit basins. Hourly streamflow observations at these gauges are obtained from the USGS REST API[63]. The CAMELS[61] dataset is used for comparison with existing hourly modeling benchmarks, including MTS-LSTM[11] and MF-LSTM[12], using the publicly released simulations from those studies.

Meteorological forcings are derived from the NOAA Analysis of Record for Calibration (AORC) dataset[48]. The hourly forcing variables used by the model are precipitation, air temperature, and potential evapotranspiration. Potential evapotranspiration is calculated from AORC meteorological fields using the Penman-Monteith formulation[64]. All forcing variables are aggregated from the original 1-km AORC grid to the Hydrofabric unit-basin scale before being passed to the runoff-generation model.

Static basin and reach attributes are used as inputs to the parameter neural networks. The runoff-generation networks use climatic, vegetation, topographic, soil, geologic, and drainage-area descriptors following the attribute set used in prior distributed differentiable HBV modeling[39]. The routing networks additionally use reach-scale attributes such as reach length, bed slope, upstream drainage area, Strahler order, upstream degree, and sinuosity, obtained or calculated from Hydrofabric. All variables used are listed in Supplementary Table S1.

*Training and evaluation*

We primarily train and evaluate δHBV2.0MTS-MC. Four ablation configurations are used to isolate the contribution of key components of the framework. First, δHBV2.0MTS-UH replaces Muskingum-Cunge river-network routing with distributed unit-hydrograph routing, thereby isolating the benefit of explicit channel routing. Second, δHBV2.0MTS-MC-base excludes the hourly fast-runoff formulation introduced above, allowing us to quantify the effects of faster saturation-excess release and infiltration-excess runoff. Third, δHBV2.0h-MC is trained independently as an hourly-only model using a 1,440-hour warm-up, providing a direct comparison with the multi-timescale warm-up; longer hourly warm-ups are substantially less efficient and exceed available GPU memory without additional engineering. Fourth, the original daily δHBV2.0-MC model identical to Song et al.[39] is used to assess the benefit of hourly modeling. In the proposed hourly configurations, runoff-generation and routing parameters are learned jointly from gauge streamflow observations.

The temporal split is fixed across experiments. The training period is 1991-01-01 to 2003-12-31, the validation period is 2004-01-01 to 2008-12-31, and the test period is 2009-01-01 to 2018-12-31. Model skill is evaluated on the independent test period using gauge-level streamflow metrics. We report NSE[65], Kling-Gupta efficiency (KGE)[66], Pearson correlation (Corr), percent bias (PBIAS), high-flow bias (FHV), and low-flow bias (FLV). FHV is computed over flows above the 98th percentile of observed streamflow, and FLV is computed over flows below the 30th percentile. For flood-event evaluation, we apply peak-over-threshold (POT) sampling[67] for gauges with at least 10 flood-sample years. At each gauge, POT3 events are defined as the largest 3*n* hourly peaks over *n* valid years, subject to a minimum inter-event separation of seven days. A generalized Pareto[68] distribution is fitted to the sampled peaks to estimate return periods. For each event, an initial window centered on the observed peak is refined to the nearest local flow minima before and after the peak, defining the event start and end; event peakiness is then calculated as the peak flow divided by the mean flow within this window, with larger values indicating sharper hydrographs. For each observed event, the corresponding simulated peak is identified within a ±24-hour window, and performance is evaluated using absolute percentage error (APE) and timing error (TE) for the hourly peak. Additional metric details are provided in Supplementary Text A.

**Data Availability Statement**

Hourly streamflow observations for the study gauges were obtained from the USGS REST API[63]. Meteorological forcings were derived from the NOAA Analysis of Record for Calibration (AORC) dataset[69], which is available through the AWS Open Data Registry (https://registry.opendata.aws/noaa-nws-aorc). Hydrofabric data[62] were obtained from the NOAA Office of Water Prediction Hydrofabric repository (https://github.com/NOAA-OWP/hydrofabric). CAMELS[70] gauge metadata are available from https://zenodo.org/records/15529996. CONUS gauge metadata are described in Song et al. [39]. MTS-LSTM simulations[71] were downloaded from https://zenodo.org/records/18953458. MF-LSTM simulations[72] were downloaded from https://zenodo.org/records/14780059. NWM3.0 simulations[73] were downloaded from

https://registry.opendata.aws/nwm-archive. LSTM-Troute simulations were generated using NOAA Office of Water Prediction's Basic Model Interface[50] (https://github.com/NOAA-OWP/lstm) and Next-generation in a box (NGIAB)[27].